# Free-running Sn precipitates: an efficient phase separation mechanism for metastable Ge$_{1-x}$Sn$_x$ epilayers

H. Groiss[1, 2, 3, 4, *], M. Glaser[1], M. Schatzl[1], M. Brehm[1], D. Gerthsen[4], D. Roth[5], P. Bauer[5], and F. Schäffler[1]

[1] Institute of Semiconductor and Solid State Physics, Johannes Kepler University Linz, Altenberger Str. 69, 4040 Linz, Austria

[2] Center of Surface and Nanoanalytics (ZONA), Johannes Kepler University Linz, Altenberger Str. 69, 4040 Linz, Austria

[3] CEST Competence Center for Electrochemical Surface Technology, Viktor Kaplan Straße 2, 2700 Wiener Neustadt, Austria

[4] Laboratory for Electron Microscopy, Karlsruhe Institute of Technology, Engesserstr. 7, 76131 Karlsruhe, Germany

[5] Institute of Experimental Physics, Division Atomic Physics and Surface Science, Johannes Kepler University Linz, Altenberger Str. 69, 4040 Linz, Austria

* corresponding author, heiko.groiss@jku.at

**Abstract**

We report on the temperature stability of pseudomorphic $Ge_{1-x}Sn_x$ films grown by molecular beam epitaxy on Ge(001) substrates. Both the growth temperature-dependence and the influence of post-growth annealing steps were investigated. In either case we observe that decomposition of metastable epilayers with Sn concentrations around 10% sets in above ≈230°C, the eutectic temperature of the Ge/Sn system. Time-resolved annealing experiments in a scanning electron microscope reveal the crucial role of liquid Sn droplets in this phase separation process. Driven by a gradient of the chemical potential, the Sn droplets move on the surface along preferential crystallographic directions, thereby taking up Sn and Ge from the strained $Ge_{1-x}Sn_x$ layer at their leading edge. While Sn-uptake increases the volume of the melt, dissolved Ge becomes re-deposited by a liquid-phase epitaxial process at the trailing edge of the droplet. Secondary droplets are launched from the rims of the single-crystalline Ge trails into intact regions of the GeSn film, leading to an avalanche-like transformation front between the GeSn film and re-deposited Ge. This process makes phase separation of metastable GeSn layers particularly efficient at rather low temperatures.

In the last few years, interest in direct-gap group-IV heterostructures has mainly been driven by the search for CMOS-compatible light emitters for monolithically integrated optical communication devices,[1] or for on-chip optical interconnects.[2,3] Sn-containing group-IV alloys are attractive because their band gap can be engineered all the way from a semiconductor to a semimetal,[4] and, moreover, they are the only known group-IV semiconductors that can assume a direct band gap.[5,6,7,8] In this respect, a seminal breakthrough was achieved with the recent demonstration of lasing in strain-relaxed $Ge_{1-x}Sn_x$ epilayers with Sn concentrations around 10%.[9,10]

Despite this major accomplishment, the thermal stability of $Ge_{1-x}Sn_x$ films with such high Sn concentrations has always been a serious concern. Sn has a lattice mismatch of 14.6% with respect to Ge concomitant with a miscibility gap over almost the entire composition range[11]. Moreover, at 13.2°C pure Sn undergoes an allotropic phase transition from the diamond lattice of semi-metallic $\alpha$-Sn to the tetragonal lattice of metallic $\beta$-Sn. Thus, a precondition for utilizing $Ge_{1-x}Sn_x$ alloys in meaningful device applications is the feasibility of metastable epitaxial growth with Sn concentrations far above the solid solubility limit of $x_{sl} \approx 1\%$[11].

Over the last 30 years a wide range of growth parameters has been investigated aiming at the implementation of metastable epitaxial films and layer sequences containing $Si_{1-y}Sn_y$ or $Ge_{1-x}Sn_x$ alloys. It could be demonstrated that far from thermal equilibrium metastable $Ge_{1-x}Sn_x$ epilayers can be grown with $x \gg x_{sl}$.[5,9,12,13,14,15,16,17,18] Still, for Si-compatible device integration the low eutectic temperature[19] of the Ge-Sn binary alloy of $T_{EC} = 231°C$[11] remains a problem.

The temperature stability of substitutionally incorporated Sn in diamond-type host lattices has been investigated by several groups.[5,16,17,20,21] In particular, Sn precipitation at the growth front of $Ge_{1-x}Sn_x$ films on Ge(001) was observed already at growth temperatures $T_G > 150°C$.[22, 23] Also, the authors of Ref. 23 were the first to report trails behind the Sn precipitates which indicate surface movement. This finding has meanwhile been confirmed by several groups, but little is known about the underlying mechanisms that propel the Sn precipitates.[24, 25, 26, 27]

Here, we report systematic growth and *in-situ* annealing experiments conducted on uncapped $Ge_{1-x}Sn_x$ films grown on Ge(001) by molecular beam epitaxy (MBE). Commercial Ge(001) substrates were employed to rule out any influence of the high threading dislocation densities and local strain variations[28] associated with virtual substrates.[29] Details of layer growth and their characterization by X-ray diffraction (XRD), Rutherford back-scattering (RBS), scanning electron microscopy (SEM), atomic force microscopy (AFM), transmission electron microscopy (TEM) and energy dispersive X-ray spectroscopy (EDXS) are described in section S1 of the Supplementary Material. The growth parameters of the four sample series A – D used in this study are listed ibidem in Table S1, and experimental details supplementing the results in the main text are given in sections S2-S6.

Sample series A was grown as a reference for $Ge_{1-x}Sn_x$ epilayers with various compositions. In brief, 30nm thick $Ge_{1-x}Sn_x$ films grown at $T_G = 120°C$ show X-ray rocking curves with very well-behaved pendellösung fringes[30] up to $x = 14\%$. In this composition range, RBS experiments reveal the validity of Vegard's law, i.e. a linear increase of the lattice constant with x (Fig. S2.1). This finding confirms the results in Ref. 16, but disagrees with experiments reported in Ref. 17, where deviations from Vegard's were claimed to set in above $x = 8\%$. TEM investigations did not show extended defects or alloy inhomogeneities up to $x = 14\%$. With higher Sn concentrations, however, the crystal quality decreases rapidly (Fig. S2.2), concomitant with the loss of pendellösung fringes in the XRD experiments.

In the following, we concentrate on systematic variations of the growth- and annealing temperatures on samples with an application-relevant Sn concentration of x = 10%. Fig. 1(a) shows X-ray rocking curves from samples of series B, for which $T_G$ was varied between 150 and 275°C. Up to $T_G$ = 225°C, i.e. just below $T_{EC}$, the pendellösung fringes are very well resolved, whereas the rocking curves of the two samples grown at 250 and 275°C show only two weak shoulders on the compressively strained side to the Ge (004) substrate peak. Evidently, the homogeneity of the GeSn layers gets lost during film growth above $T_{EC}$.

To further assess this finding, we recorded AFM, SEM and TEM images of the degraded samples of series B (Fig. S3). These samples show a rough surface with partly embedded droplet-shaped objects. TEM and EDXS investigations revealed that the film between the droplets consists of single crystalline Ge with a small fraction of dissolved Sn, whereas the (solidified) droplets consist of β-Sn. Thus, above $T_{EC}$ the $Ge_{0.9}Sn_{0.1}$ films become phase-separated already during MBE growth, with the Sn phase segregating at the film surface in liquid form as inferred from the shape of the precipitates.

A similar temperature dependence was observed on the sample from series C, for which $Ge_{0.9}Sn_{0.1}$ films were grown in a fully coherent manner at $T_G$ = 200°C, i.e. below $T_{EC}$. The films were then annealed *in-situ* for 15 min at temperatures $T_A$ between 200°C and 350°C. As expected, the films remained stable at $T_A$ = 200°C, but higher annealing temperatures led to a complete loss of the pendellösung fringes and the appearance of two weak shoulders below the substrate peak (Fig. 1(b)). The SEM images in Fig. 2 depict the morphological changes that occur during annealing above $T_{EC}$. Again, droplet-shaped precipitates of varying diameters appear on the surfaces. Distinct trails appear in connection with the largest droplets which are predominantly oriented along the <110> directions of the substrate. Movement in the <100> directions is also observed, as well as transitions from one preferred direction class to the other (Fig. 2(a)). SEM images with higher magnification (Figs. 2(b) and 2(c)), and AFM images (Fig. S4) show the complex fine structure of the trails. The most prominent features are bundles of herringbone-like lines that are speckled with small Sn precipitates (Fig. 2(c), white arrows). AFM line scans revealed that the trails have essentially the same thickness as the original epilayer, except for two ≈ 60nm deep trenches that confine them laterally. Maps of the local inclination angles (Fig. S4) reveal that the herringbone pattern consists mainly of well-known low-energy facets of Ge, namely {001}, {105} and {113}[31]. Between the larger Sn droplets and their trails also smaller precipitates are present which come with their own, more unsteady trails. Only small patches remain smooth, and therefore seem to be un-affected by the phase separation process (Fig. 2(c), black arrows).

To assess the decomposition mechanism in more detail, we performed post-growth annealing experiments in the high-vacuum environment of a SEM instrument with a temperature-controlled sample holder. For this purpose, samples from series D were transferred from the MBE chamber into the SEM under ambient conditions in less than five minutes to minimize surface reactions with the atmosphere. Secondary electron images were taken simultaneously with an Everhart-Thornley (SE2)[32] and a GEMINI In-lens detector. The former shows a combination of surface topography and material contrast[32], whereas the latter is only sensitive to the topography. To visualize the dynamics of droplet movement, we compiled image sequences to four stop-motion video clips (V1 – V4) which are available in Section S5 of the Supplementary Materials. V1 – V3 were recorded near $T_{EC}$ at 250°C, V4 at 350°C, with estimated accuracies of ±25°C.

The video sequences clearly show that each Sn droplet defines the very location of a transformation process that converts the intact $Ge_{1-x}Sn_x$ film at its leading edge into liquid Sn, which is increasing the droplet volume, and a corrugated trail region. The droplets come to a halt as they run into trail regions of other droplets. This can be seen in Figs. 2(d)-(e) which are extracted from video clip V1. This video sequence follows the movement of the central one of three large droplets (marked with a white arrow in Fig. 2(d)), until it is stopped by the trail of a droplet crossing its path (Figs. 2(e) and (f)). Evidently, the Sn precipitates can only move if they are simultaneously in contact with an intact region of the GeSn layer and the corrugated trail region. From the large trails smaller, oblate droplets start to move into intact regions of the GeSn film. These behave in a similar way as the large droplets, thus carrying the transformation process in an avalanche-like manner into areas between the trails of large dots. Video sequences V2 and V3 show additional experiments with further details of the phase transformation process. V4 was recorded at $T_A = 350°C$ where a higher density of Sn droplets is observed. V4 demonstrates the remarkable efficiency of the underlying transformation mechanism at a temperature that is still moderate in comparison with typical CMOS processing temperatures.

To gain quantitative information on the role of the molten Sn precipitates in the transformation process, we performed TEM experiments on samples from Series D after annealing. For this purpose, a 20 µm long TEM-lamella was cut with a focused ion-beam (FIB) through a large Sn droplet and its surroundings. A pair of electron transparent windows was then prepared in regions ahead and behind the Sn droplet (Figs. 3(a), (b)) to determine the local compositions and strains in these regions. The high-resolution TEM images in Figs. 3(c) and (e) reveal excellent crystal quality in either region. Employing EDXS we found that the trail region consists of almost pure

Ge with x ≤ 1%. Only the topmost few monolayers of the film contain segregated Sn. The corresponding lattice constants were extracted by Fourier transformations (FFT) of cross-sectional areas that contain both the Ge substrate and the annealed epilayer. As a result, we identified the trail region to consist of virtually strain-free Ge, whereas the GeSn film ahead of the Sn droplet preserved its original strain and composition (Fig. 3(d) and (f)). The TEM images also revealed that the small Sn droplets in the trail (white arrows in Fig. 2(c)) decorate {111}-faceted pits[31]. In a final experiment, we show with cross-sectional TEM-lamellae through large Sn droplets that the large precipitates extend down to the interface with the Ge buffer (Fig. 4(a)). We also used such a specimen to determine the crystal structures and -orientations of the different phases after cool-down (Supplementary Materials section S6).

Our experiments showed that above the eutectic temperature free-running liquid Sn droplets induce the phase separation of a strained GeSn film into liquid Sn and crystalline Ge with negligible Sn content. This process is schematically illustrated in Fig. 4(b). In this picture, Ge is removed from the leading edge of the droplet, transported through the melt and re-deposited at the trailing edge. It is this directional flow of Ge that causes droplet movement, while simultaneously the dissolved volume of the strained and metastable $Si_{1-x}Ge_x$ layer becomes converted into unstrained, crystalline Ge and liquid Sn.

This overall process differs substantially from phase separation of immiscible solids based on solid-state diffusion, as e.g., the precipitation of Si in the Si/Al system [33] or surface-mediated growth and topological transitions of nanostructures formed by immiscible phases[34, 35]. The movement of a Sn droplet resembles more the process of a free-running n-alkane droplet containing surface-active agents. If the latter form grafted hydrophobic layers on an originally hydrophilic surface, droplet movement into hydrophilic areas becomes initiated.[36] In our case the phase separation process itself is much more complex and reminiscent of liquid phase epitaxy (LPE) in which a precursor species is dissolved in a supersaturated melt from where it precipitates epitaxially when brought in contact with a crystalline substrate.[37] In conventional, homoepitaxial LPE, feeding of the melt with the precursor and epitaxial growth are two separate processes. These have to be conducted at different temperatures in order to achieve a supersaturated melt. In our particular case, feeding and growth happen simultaneously in every free-running Sn droplet, as long as it stays in contact with both the SnGe layer and the Ge trail.

In analogy to conventional LPE, epitaxial Ge growth at the trailing edge of each droplet is described by a negative difference of the Gibbs free energies of melt and crystalline Ge, $\Delta G^{Ge}$.[38, 39]

$$\Delta G^{Ge} = (\mu_s^{Ge} - \mu_l) + \Delta\gamma_{ls}^{Ge} < 0. \qquad \text{equ. 1}$$

Here, $(\mu_s^{Ge} - \mu_l)$ is the difference between the chemical potentials of Ge in the solid trail ($\mu_s^{Ge}$) and in the melt ($\mu_l$); $\Delta\gamma_{ls}^{Ge}$ accounts for interface energy differences induced by changes of the liquid-solid interface during growth. The observed low-energy facets in the herringbone pattern of the deposited Ge trails result from a minimization of the $\Delta\gamma_{ls}^{Ge}$ term. Under these conditions LPE growth is essentially defined by $(\mu_s^{Ge} - \mu_l) < 0$, which describes the aforementioned supersaturation of the melt with Ge.[38]

The feeding part of the LPE process occurs at the leading edge of each Sn droplet, where it is in contact with the strained GeSn film. This contact region is described by $\Delta G^{GeSn}$, which contains additional terms that account for the strained GeSn heterostructure. $\Delta G^{GeSn}$ has to be positive, because here the GeSn film dissolves in the melt. Overall, we get:[40]

$$\Delta G^{GeSn} = (\mu_s^{GeSn} - \mu_l) + (\Delta\gamma_{ls}^{GeSn} + E_{ls} + \gamma_{ss}) > 0. \qquad \text{equ. 2}$$

$\mu_s^{GeSn}$ is the chemical potential of the solid GeSn film; $E_{ls}$ accounts for the strain energy of the liquid-solid interface and $\gamma_{ss}$ for the solid-solid interface energy between the strained GeSn film and the Ge substrate.[40] Also, $\Delta\gamma_{ls}^{GeSn}$ stands for changes in the liquid-solid interface energy. Minimization of this term leads to faceting of the dissolving front, as can be seen nicely in video clip V3.

Feeding and growth are coupled in the droplet by $\mu_l$, the chemical potential of solved Ge in the liquid Sn-melt. If we assume that Ge diffusion in the droplet is much faster than the growth and dissolution kinetics, $\mu_l$ becomes approximately constant over the melt. $\mu_l$ is determined by two effects: For one, it depends on the strain- and interface terms in the second pair of parentheses in equ. 2, which becomes a minimum if, as we have observed in the experiments, the whole thickness of the epilayer is dissolved down to the substrate. On the other hand $\mu_l$ is determined by the phase separation process itself. Since Ge and Sn are essentially immiscible, their separation into pure Ge and a Sn melt is energetically favorable, releasing essentially the mixing enthalpy $\Delta H_{mix}^{GeSn}$ of the Ge$_{1-x}$Sn$_x$ layer. However, producing a Sn melt supersaturated with Ge costs energy. This we express by $\Delta G_l^{Ge}$ which is a function of the Ge concentration (and thus of $\mu_l$) in the melt. This leads then to a change of the Gibbs free energy associated with the phase separation process of:

$$\Delta G^{Ge+Sn(l)} = \Delta G_l^{Ge} - \Delta H_{mix}^{GeSn} \qquad \text{equ. 3}$$

A more detailed discussion of equ. 3 can be found in the Supplementary Material S7. The contributions $\mu_s^{GeSn}$ and $\Delta H_{mix}^{GeSn}$ in equ. 2 and 3 are only known for equilibrium processes,[11, 41, 42, 43] but not for our case of far-from-equilibrium MBE-growth. It is, however, clear from the experimental observations that the system gains free energy when Ge and Sn from the metastable GeSn film become dissolved in the Sn melt, and simultaneously almost pure (x ≤ 1%)[11] Ge is deposited epitaxially at the opposite side of the melt.

To estimate the steady-state Ge concentration in the larger Sn droplets during their movement, we evaluated TEM images after cool-down. During solidification, the Ge-content in the Sn droplet is reduced to the equilibrium solubility at the melting point.[11] This effect leads to precipitation of the excess Ge content in the shape of a collar around the Sn-droplet, as can be seen in Figs. 2(b) and S5. An estimate of the collar's volume in respect to the volume of the droplet led us to the conclusion that more than 11% Ge must have been dissolved in the liquid droplet. This value is much higher than the equilibrium solubility of 2-3% in the investigated temperature window between $T_{EC}$ and 350°C.[11] Evidently, the dissolution of Sn and Ge from the strained and metastable SnGe film leads to a high degree of supersaturation in the melt, which allows for Ge epitaxy at the trailing edge.

Efficient phase separation at such low temperatures imposes severe limitation to applications based on metastable GeSn films. It is therefore necessary to suppress Sn precipitation or at least to shift it to higher processing temperatures. Several such measures are conceivable: (i) Capping of the GeSn epilayers affects Sn diffusion in the solid phase and thus delays the formation of sufficiently large Sn precipitates that are required to initiate the transformation process. (ii) Rapid thermal annealing, which is routinely applied in high-temperature CMOS processes, limits the amount of segregated Sn,[29] as compared to the long-term, quasi-equilibrium annealing conditions in our experiments. (iii) Growth by CVD under kinetic conditions very far from equilibrium has been shown to suppress Sn precipitation during growth up to temperatures of at least 390°C.[9] The extension of the metastable range to temperatures substantially above $T_{EC}$ is most likely based on two beneficial properties of the particular CVD process utilized in Ref. 9: For one, hydrogen termination of the surface during CVD is expected to effectively suppress surface segregation of Sn, and thus the initial source of molten Sn during growth. Secondly, suppressed Sn segregation and the far-from-equilibrium growth conditions allow for higher growth temperatures, thus reducing the density of point defects that may play a role for solid-state Sn diffusion in our low-temperature MBE material.

It remains to be seen, to what extent the expanded temperature window for GeSn growth by CVD will allow sufficiently high processing temperatures for device integration. For now, the results in Ref. 24 gained from such materials suggest that Sn precipitation can be delayed to higher processing temperatures, but not totally suppressed.

In summary, we investigated the thermal stability of uncapped $Ge_{0.9}Sn_{0.1}$ films grown by MBE on Ge(001) substrates. Above the eutectic temperature of 231°C, we find an efficient phase separation mechanism based on molten Sn precipitates that move over the surface. The free-running Sn droplets induce phase separation by taking up Sn and Ge from the intact GeSn film at their leading, and precipitating crystalline Ge at their trailing edges. This behavior is attributed to a liquid-phase epitaxial process that is driven by the free-energy difference between the GeSn and the Ge layer which are both in contact with the molten Sn droplet during movement.

Supplementary Material

Please see the Supplementary Materials for detailed information of the layer growth and characterization methods (section S1). Additional experimental results and descriptions of the supplementary video files (V1-V4) can be found in the sections S2-S6. Details of the thermodynamic model are presented in section S7.

13Acknowledgments

This work was financially supported by the Austrian Science Fund (FWF Vienna, Austria), via projects SFB IRON (F2502-N17), J-3317, and P29137-N36. Financial support within the Comet Program of the Austrian Research Promotion Agency (FFG) via Grant 844596 and by the Governments of Lower and Upper Austria are gratefully acknowledged.

## Acknowledgments

This work was financially supported by the Austrian Science Fund (FWF Vienna, Austria), via projects SFB IRON (F2502-N17), J-3317, and P29137-N36. Financial support within the Comet Program of the Austrian Research Promotion Agency (FFG) via Grant 844596 and by the Governments of Lower and Upper Austria are gratefully acknowledged.

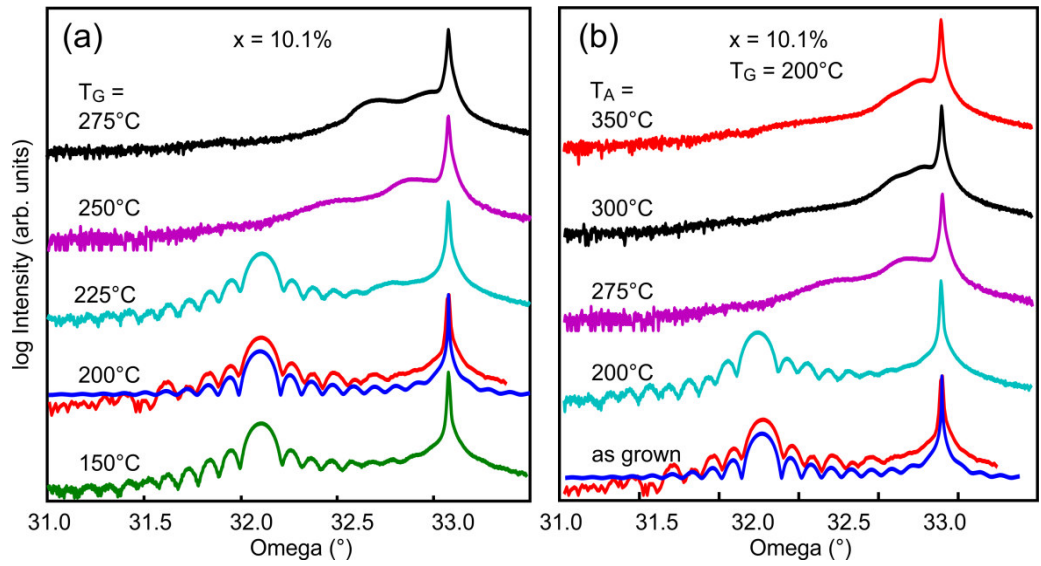

Fig. 1: (a) X-ray rocking curves from samples of Series B, which were grown at increasing growth temperature $T_G$. (b) Rocking curves from samples of Series C, which were annealed *in-situ* at increasing annealing temperatures $T_A$.

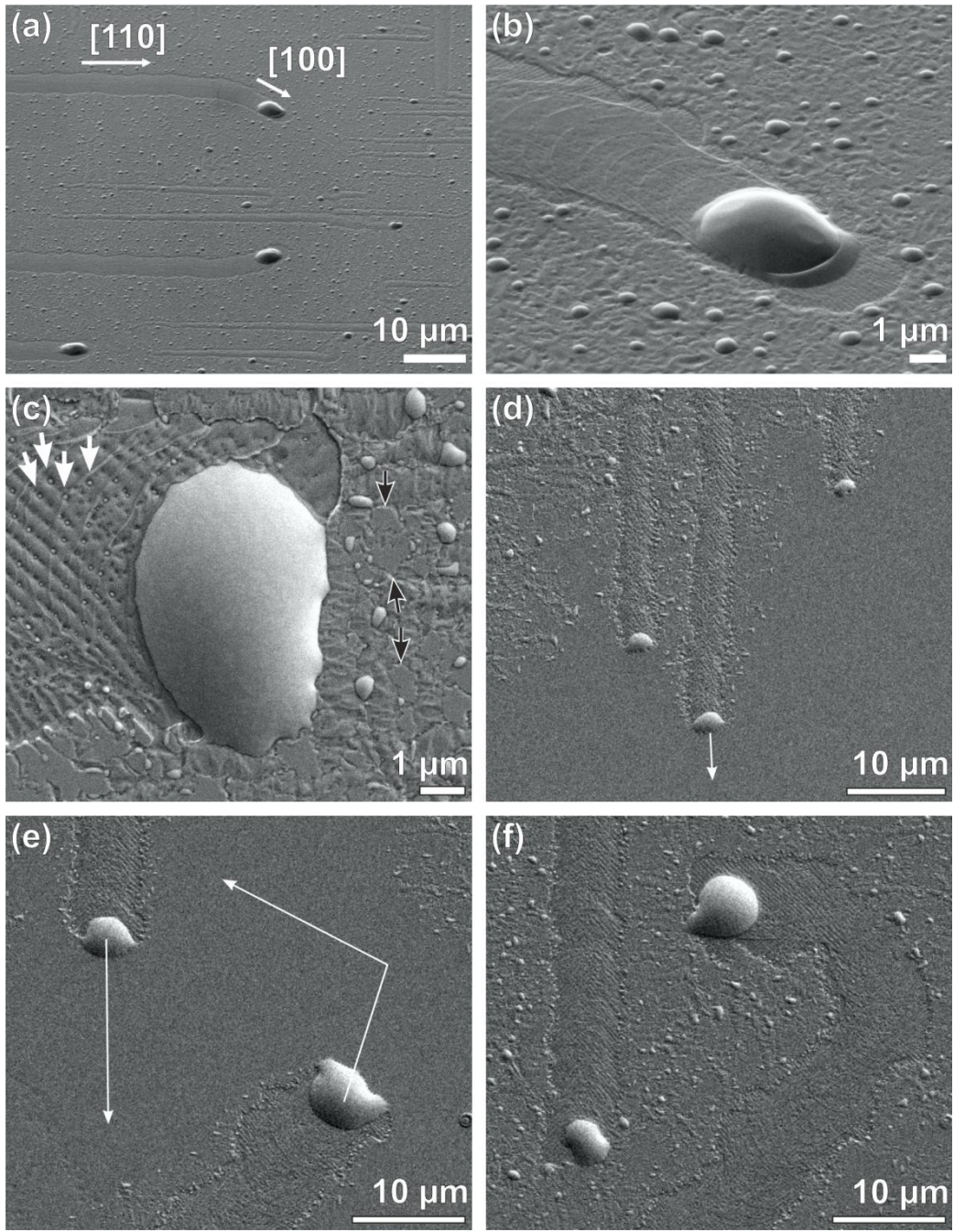

Fig 2: (a) and (b) SEM overviews of sample C4 after annealing and cool-down. The trajectories of the Sn droplets follow either <110> or <100> directions. (c) Magnified SEM image of a large Sn droplet after *in-situ* annealing. The black arrows mark residual areas of the intact $Ge_{1-x}Sn_x$ layer. The white arrows point at some of the many tiny Sn droplets that decorate the trail. The complex pattern of the trail containing mainly {001}, {105} and {113} facets is well visible. (d)-(f) Still images from video clip V1 following the central droplet marked with a white arrow. (e) shows the near-by crossing of a droplet shortly before both movement comes to a halt, which is displayed in the follow-up image (f).

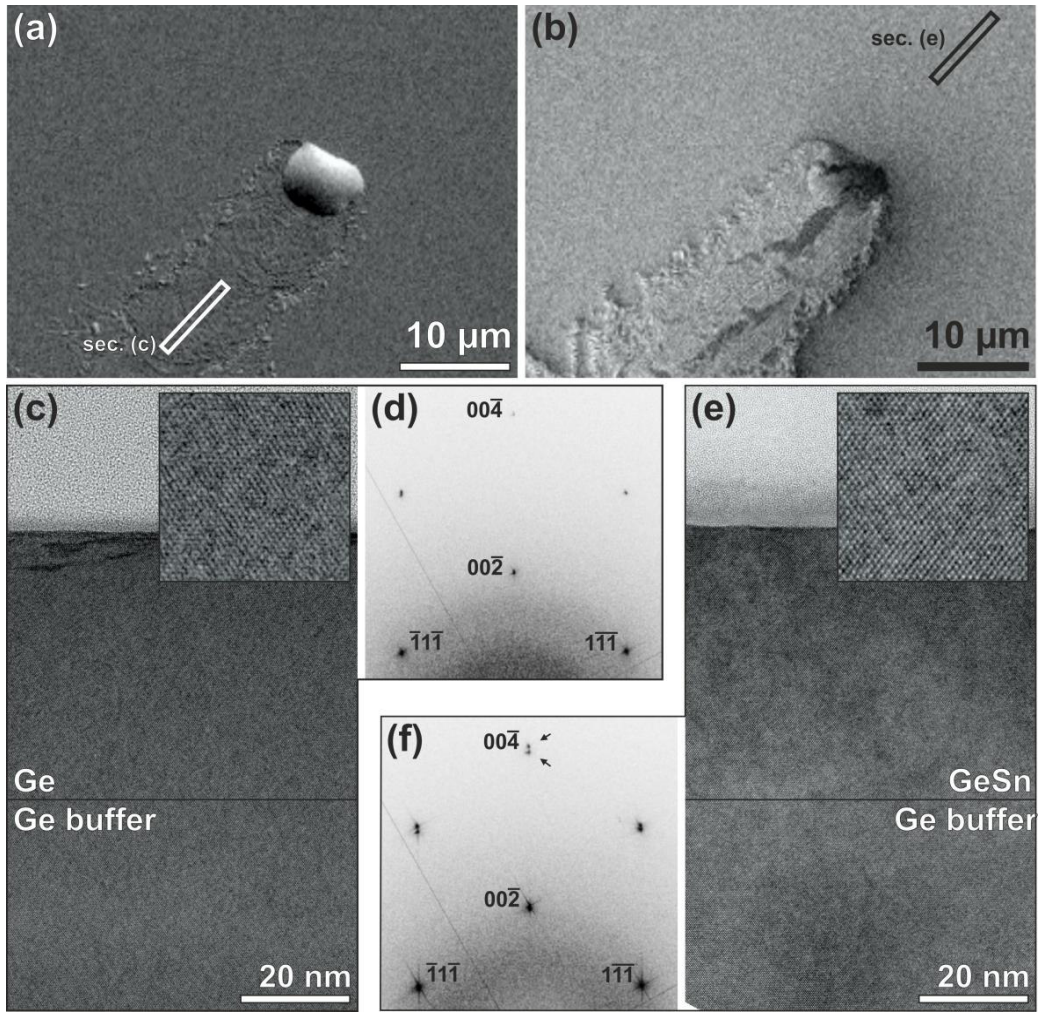

Fig. 3: SE2 (a) and In-Lens (b) images of a large Sn droplet at the phase separation front. (c) HRTEM cross-sectional images from the trail region marked in (a). The calculated diffraction pattern (FFT) of (c) is displayed in (d). (e) and (f) show the HRTEM cross-sectional image and the extracted diffraction pattern from the strained GeSn region ahead of the Sn droplet, as indicated in (b).

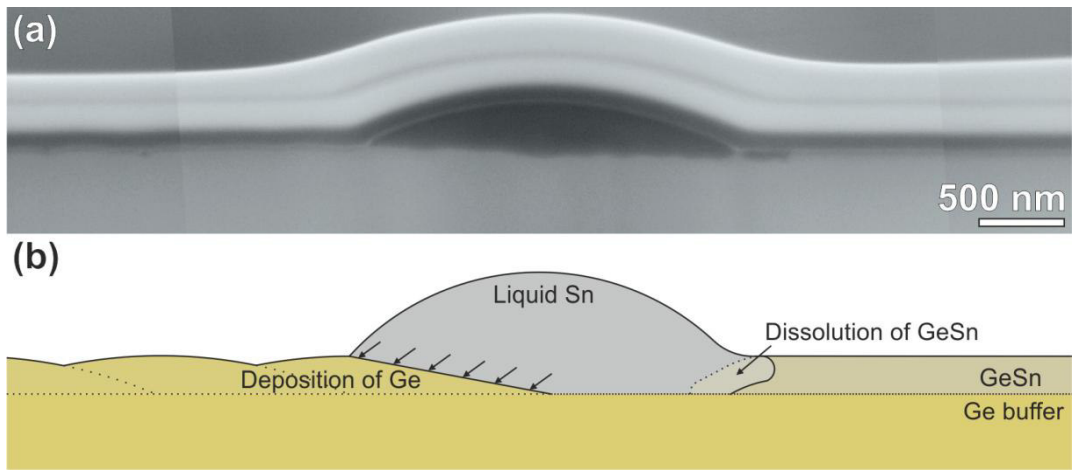

Fig. 4: (a) Cross-sectional SEM image of a Sn droplet that separates the Ge epilayer on the left side from the intact GeSn film on the right side. To protect the solidified droplet during preparation the sample was covered with e-beam and FIB-induced Pt-depositions. (b) Schematic view of the phase separation process induced by the liquid Sn droplet.

Supplementary Materials for:

Free-running Sn precipitates: an efficient phase separation mechanism for metastable $Ge_{1-x}Sn_x$ epilayers

H. Groiss[1, 2, 3, 4, *], M. Glaser[1], M. Schatzl[1], M. Brehm[1], D. Gerthsen[4], D. Roth[5], P. Bauer[5], and F. Schäffler[1]

[1] Institute of Semiconductor and Solid State Physics, Johannes Kepler University Linz, Altenberger Str. 69, 4040 Linz, Austria

[2] Center of Surface and Nanoanalytics (ZONA), Johannes Kepler University Linz, Altenberger Str. 69, 4040 Linz, Austria

[3] CEST Competence Center for Electrochemical Surface Technology, Viktor Kaplan Straße 2, 2700 Wiener Neustadt, Austria

[4] Laboratory for Electron Microscopy, Karlsruhe Institute of Technology, Engesserstr. 7, 76131 Karlsruhe, Germany

[5] Institute of Experimental Physics, Division Atomic Physics and Surface Science, Johannes Kepler University Linz, Altenberger Str. 69, 4040 Linz, Austria

[*] corresponding author, heiko.groiss@jku.at

**Supplementary Material S1 – Experimental Section**

**Epitaxial Growth**

All samples were grown in a Riber Siva 45 MBE facility with electron-beam evaporators for Si and Ge. For this work, we installed an additional effusion cell for high-purity Sn which was calibrated by secondary-ion mass spectrometry (SIMS) of GeSn superlattices in an analogous way as described in Ref. 1. The samples are heated radiatively with temperature control calibrated to within ±25°C. Polished Ge (001) substrates with a diameter of 100 mm and a specified resistivity of 8 Ωcm were purchased from *Umicore* and subsequently diced into 9.5×9.5 mm² pieces to fit into solder-free adapters milled from high-purity Si ingots[2]. These commercial Ge(001) substrates were employed to rule out any influence of the high threading dislocation densities and local strain variations typically associated with virtual Ge[3] (also called: Ge-buffered[4]) substrates. The Ge substrate pieces were chemically pre-cleaned[5] immediately before being introduced into the load-lock chamber of the MBE system. Before growth, the Ge-substrates were degassed for 30 min at 300°C and then heated for 15 min to 750°C for oxide desorption. Growth always commenced with a 50 nm thick Ge buffer layer deposited at 400°C which results in smooth surfaces with double-atomic height steps only[2]. The substrate temperature was then ramped down to the respective growth temperature $T_G$ of the $Ge_{1-x}Sn_x$ epilayer.

To assess the thermal stability and the precipitation kinetics of $Ge_{1-x}Sn_x$ epilayers under systematic and well-controlled experimental conditions we grew four series of un-capped $Ge_{1-x}Sn_x$ films (Table S1). Most of the samples of Series A were grown at a low temperature of $T_G = 120°C$ to calibrated our sources and growth parameters, and to assess the composition range x in which we can achieve coherent growth of metastable $Ge_{1-x}Sn_x$ without precipitation. In addition, we also grew two samples (A8 and A9) at $T_G = 200°C$ and at a higher growth rate to study the influence of growth temperatures slightly below the eutectic temperature of the Ge/Sn system. All samples of Series A were characterized by X-ray diffraction (XRD) and atomic force microscopy (AFM). Samples A1 – A4 and A9 were used for Rutherford Backscattering (RBS) experiments, and samples A7 - A9 were investigated by transmission electron microscopy (TEM).

In Series B to D we concentrated on an application-relevant[4] Sn concentration of x = 10% and performed different temperature stability experiments. In Series B we increased systematically the growth temperature $T_G$, whereas in Series C films grown at $T_G = 200°C$ were *in-situ* annealed in the ultra-high vacuum environment of the MBE chamber for 15 min at increasing temperatures. Finally, the annealing step on the samples of Series D was performed *ex-situ* in the high-vacuum environment of a LEO Supra 35 scanning electron microscope (SEM) with a heatable sample stage. Transfer from the MBE chamber to the SEM occurred under ambient conditions in less than five minutes to minimize surface reactions with the atmosphere. Film compositions, growth rates and film thicknesses of the four investigated sample series are listed in Table S1.

| sample # | $T_G$ (°C) | $T_A$ (°C) | $x_{RBS}$ (%) | $x_{XRD}$ (%) | d (nm) | r (nm/s) |
|---|---|---|---|---|---|---|
| Series A | | | | | | |
| A1 | 120 | — | 5.13 | 4.9 | 30 | 0.015 |
| A2 | 120 | — | 6.70 | 7.3 | 30 | 0.015 |
| A3 | 120 | — | 9.25 | 9.4 | 30 | 0.015 |
| A4 | 120 | — | 12.70 | 12.8 | 30 | 0.015 |
| A5 | 120 | — | — | 13.6 | 30 | 0.015 |
| A6 | 120 | — | — | 14.5 | 30 | 0.015 |
| A7 | 120 | — | — | 15.0 | 30 | 0.005 |
| A8 | 120 | — | — | 5.0 | 100 | 0.015 |
| A9 | 200 | | 11.00 | 11.5 | 100 | 0.1 |
| A10 | 200 | — | — | 8.2 | 250 | 0.1 |
| | | | | | | |
| Series B | | | | | | |
| B1 | 150 | — | — | 10.1 | 50 | 0.1 |
| B2 | 200 | — | — | 10.1 | 50 | 0.1 |
| B3 | 225 | — | — | — | 50 | 0.1 |
| B4 | 250 | — | — | — | 50 | 0.1 |
| B5 | 275 | — | — | — | 50 | 0.1 |
| B6 | 300 | — | — | — | 50 | 0.1 |
| Series C | | | | | | |
| C1 = B2 | 200 | — | — | 10.1 | 50 | 0.1 |
| C2 | 200 | 200 | — | 10.1 | 50 | 0.1 |
| C3 | 200 | 275 | — | — | 50 | 0.1 |
| C4 | 200 | 300 | — | — | 50 | 0.1 |
| C5 | 200 | 350 | — | — | 50 | 0.1 |
| Series D | | | | | | |
| D1 | 200 | ≥ 230 | | 10.1 | 50 | 0.1 |
| D2 | 200 | ≥ 230 | | 10.1 | 50 | 0.1 |

**Table S1**: Growth and annealing parameters of the investigated samples. $T_G$: growth temperature; $T_A$: annealing temperature for a 15 min *in-situ* annealing step in Series C, and *ex-situ*, real time annealing in Series D, respectively; $x_{RBS}$: Sn concentration determined by Rutherford backscattering (RBS); $x_{XRD}$: Sn concentration determined by X-ray diffraction (XRD) assuming Vegard's law; d: thickness of the $Ge_{1-x}Sn_x$ film; r: deposition rate.

**X-Ray Diffraction (XRD)**

All samples were characterized by XRD on either a Seifert XRD 3003 or a PANalytical X'Pert MRD XL diffractometer, both equipped with line detectors. Routinely, rocking curves (ω-2θ scans)[6] were recorded to determine the out-of-plane lattice constants of the grown GeSn films.

**Rutherford Backscattering (RBS)**

Since XRD can only provide the lattice constants of strained epitaxial films, we also performed RBS experiments on samples A1 – A4 and A9 to assess the composition of the respective films. By comparison with the XRD experiments we then determined the relation between composition and lattice constant (Fig. S2.1 in section S2 below). The RBS measurements were conducted at the Atomic-Physics and Surface-Science Division at Johannes Kepler University in a high vacuum chamber with a base pressure in the $10^{-7}$ mbar range that is attached to an AN-700 van de Graaf accelerator. The chamber is equipped with two semiconductor surface barrier (SSB) detectors, namely a $LN_2$-cooled high resolution detector[7] featuring $\approx$3 keV full-width-at-half-maximum (FWHM) for protons and $\approx$7 keV for helium ions (scattering angle 150.1°, Cornell geometry), and a standard SSB detector of larger solid angle (scattering angle 154.6°, IBM geometry). Energy spectra of the samples were recorded using 550 keV $He^+$ ions. To avoid channeling effects and to optimize depth resolution, two angles of incidence of the ion beam ($\alpha = 0°$ and $\alpha = 60°$) were chosen. The respective Sn contents were deduced from simulations of the experimental spectra employing the SIMNRA simulation software[8].

**Scanning Electron Microscopy (SEM)**

For investigations on the Sn precipitation kinetics, we constructed a heatable sample holder with built-in temperature sensor to fit into a ZEISS Leo Supra 35 scanning electron microscope (SEM). This field-emitter SEM is equipped with a GEMINI column that allows for In-lens (inLens) detection of so-called SE1 secondary electrons which are predominantly generated by the incident electron beam.[9] In addition, a conventional Everhart-Thornley (SE2) detector is available, which provides a higher degree of material contrast due to a higher sensitivity to SE2 and SE3 electrons that are generated by back-scattered electrons on the sample surface and at structural components of the SEM instrument, respectively.[9]

**Atomic Force Microscopy (AFM)**

A Digital Instruments Veeco Dimension 3100 atomic force microscope (AFM) was used in a non-contact tapping mode to assess both the surface roughness and height profiles of the trails left behind by Sn precipitates moving on the surface. Either MICRON or Olympus TESP cantilevers were employed for this purpose. Height images and surface angle plots[10] were extracted from the AFM raw data with the free Gwyddion analysis software[11].

**Transmission Electron Microscopy (TEM)**

Transmission electron microscopy (TEM) was performed either at the Karlsruhe Institute of Technology (KIT, Karlsruhe, Germany), with a FEI TITAN[3] 80–300 at 300 kV, or in Linz with a JEOL JEM-2200FS at 200kV. Specimens were either prepared by conventional dimple grinding and subsequent argon sputtering in Karlsruhe, or in Linz with the focused ion beam (FIB) technique using a ZEISS 1540XB Cross-Beam system. The FEI TITAN[3] is equipped with an image aberration corrector, which was used for high-resolution (HR)TEM investigations. Also, scanning TEM (STEM) experiments were performed with the FEI TITAN[3] in combination with energy dispersive X-ray spectroscopy (EDXS) for composition mappings. The annealed samples were investigated with the JEM-2200FS using both HRTEM imaging and STEM-EDXS.

**Supplementary Material S2 – Experimental Results of Reference Series A**

Reference samples (Series A in Table S1) were used to calibrate our sources and growth parameters and to assess the composition range x in which we can achieve coherent growth of metastable $Ge_{1-x}Sn_x$ films without Sn precipitation. The results are summarized in Fig. S2.1 for 30nm thick $Ge_{1-x}Sn_x$ films in the range 4.9% ≤ x ≤ 14.5%, as listed in Table S1 above. A low growth temperature of $T_G$ = 120°C was chosen for most samples, but we confirmed that with an increased deposition rate virtually identical results can be achieved at $T_G$ = 200°C (samples A9, A10 in Table S1). Fig. S2.1(a) displays experimental X-ray rocking curves of the out-of-plane (004) reflex together with pendellösung simulations[6] that contain the film thickness and the lattice constant of the respective $Ge_{1-x}Sn_x$ film as the only adjustable parameters. Up to a Sn concentration of 13.6% the rocking curves show very well behaved pendellösung fringes of the thin, compressively strained $Ge_{1-x}Sn_x$ films. Above $x \approx 14\%$ the rocking curve still shows a distinct peak related to strained $Ge_{1-x}Sn_x$, but the thickness interference fringes are almost completely gone.

To assess the chemical composition of the films, Rutherford Backscatter (RBS) experiments were performed on samples A1 – A4 and A9. Fig. S2.1(b) shows, as a representative example, the RBS measurement of sample A9 together with the simulation curves for an absolute Sn concentration of 11±1%. The experiments were repeated under different incidence angles of the 550 keV He$^+$ ions to rule out channeling effects.

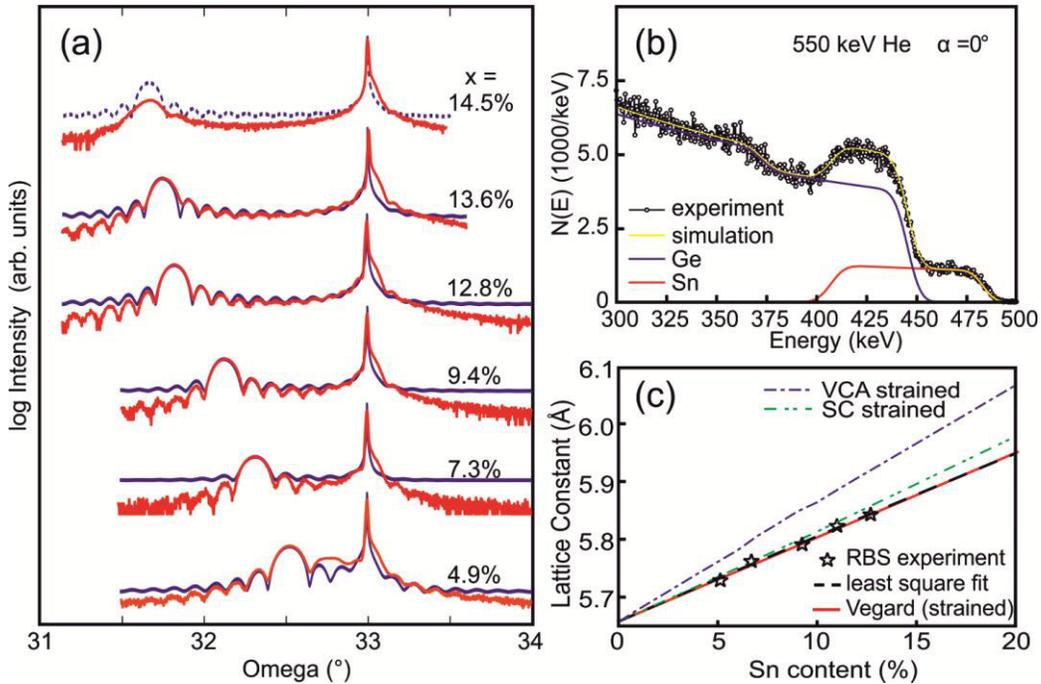

**Fig. S2.1**: (a) X-ray rocking curves of GeSn layers with increasing Sn content. (b) Representative RBS experiment on sample A9 (Table S1) and corresponding simulation for x = 11±1%. (c) Lattice constant of $Ge_{1-x}Sn_x$ with respect to the Sn content. The red line assumes a linear variation of the lattice constant with composition (Vegard's law), which coincides with the black dashed line representing a least-square fit to the RBS data points. The dash-dotted curves are taken from two recent theoretical models discussed in Ref. 14.

From the results of the X-ray rocking curves and a set of five RBS measurements we extracted the relation between the out-of-plane lattice constant and the composition of the film. The experimental data points are plotted in Fig. S2.1(c) together with a least square fit to the experimental data (black dashed line in Fig. S2.1c). Also, Vegard's law, which assumes a linear relation between lattice constant and composition, is plotted as a red line in Fig. S2.1(c). The least-square fit is almost indistinguishable from Vegard's law, which demonstrates its applicability up to $x = 14\%$ in agreement with the results in Ref. 12. We did not find any indications for the deviations from Vegard's law above $x = 8\%$, which were claimed in Ref. 13. Also, we cannot confirm the rather large bowing parameters predicted by recent density functional simulations[14] which are also depicted in Fig. S2.1(c) as reference lines.

Based on the results from Series A, the compositions of all subsequently analyzed samples were derived from X-ray diffraction experiments under the assumption of Vegard's law. Sample Series A demonstrates that MBE growth at low enough $T_G$ leads to high-quality $Ge_{1-x}Sn_x$ films up to Sn concentrations of $\approx 14\%$, thus confirming similar findings of other groups[15, 16].

Various samples from the calibration series A were investigated with transmission electron microscopy (TEM) either with STEM imaging (with bright field (BF) and high-angle annular dark field (HAADF) detectors) or HRTEM in parallel illumination mode. We use HRTEM of cross-section specimens to evaluate the crystal quality. Local lattice parameters were determined from reciprocal space images generated by Fast Fourier Transformation (FFT). Figs. S2.2(a)-(c) present TEM results of sample A5 with $x = 0.136$, and Figs. 2.2(d)-(f) of sample A7 with $x = 0.15$. Figs. S2.2(a) and (d) show BF images of the two layers. In sample A5 we found a continuous layer of high crystal quality, as can be seen in the corresponding HRTEM image of Fig. S2.2(b). The sample with 15% Sn contains extended defects (marked with black arrows in Fig. S2.2 (d)), which reach from the surface to the $Ge_{1-x}Sn_x/Ge$ interface. The HRTEM image in Fig. 2.2(e) shows that the crystal quality is still high in the vicinity of the defect. The quality of the lattice fringes decreases at defects (marked with a white arrow), indicating a heavily distorted lattice that might be induced by strong relaxation effects or, perhaps, by Sn interstitials. Figs. S2.2(c) and (f) show the corresponding reciprocal space images generated by FFT from HRTEM images that contain information from both the GeSn layer and the Ge substrate. Only a common *fcc*-based pattern (diamond structure) in $\langle 110 \rangle$ direction is visible, indicating that no precipitates with different crystal structure are present. The two zoom-ins in Figs. S2.2(c) and (f) show the $(006)$ and $(4\bar{4}0)$ reflexes, which are assigned by labels to their corresponding origin in the material stack. The $Ge_{1-x}Sn_x$ layer with $x = 13.6\%$ is coherently strained (tetragonally distorted), which leads to the coincidence of the oblique $(4\bar{4}0)$ spots of the Ge buffer and the GeSn layer.

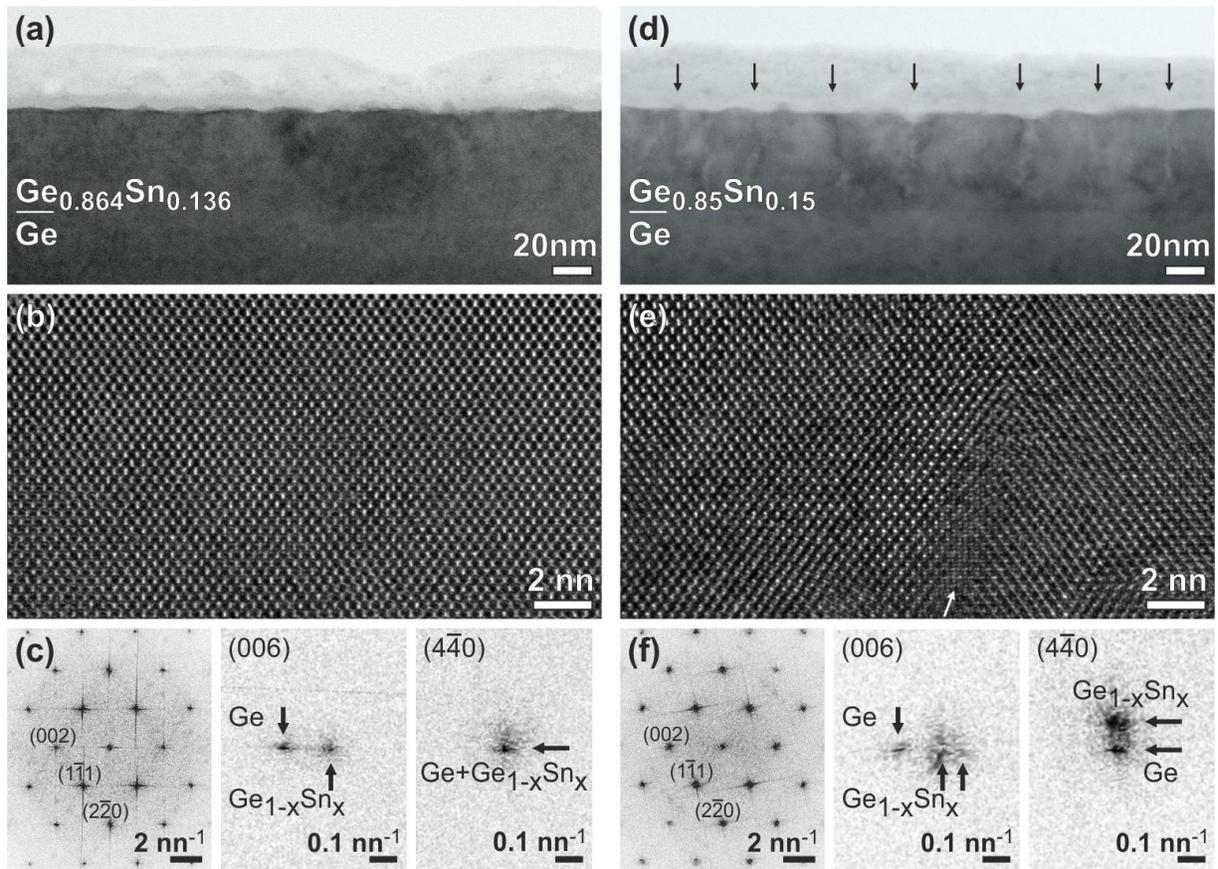

**Fig. S2.2**: (a) and (d) show larger-area BF TEM images of sample A5 (x = 0.136) and A7 (x = 0.15), (b) and (e) the corresponding HRTEM images. With increasing Sn content crystal inhomogeneities and blurred lattice fringes appear (white arrow in (e)). (c) and (f): Corresponding reciprocal space images and zoom-ins of the $(006)$ and the $(4\bar{4}0)$ reflexes generated by FFT from HRTEM images that contain both the GeSn film and the Ge substrate.

This is not the case for the GeSn film with 15% Sn. Here, the $(4\bar{4}0)$ reflex splits, which is indicative of (partial) relaxation of the GeSn film. Lattice relaxation is not homogeneous, as can be inferred from the fact that the $(006)$ reflex associated with the $Ge_{1-x}Sn_x$ layer splits. Thus, under the growth conditions used for Series A, a limit for the substitutional incorporation of Sn into a tetragonally strained $Ge_{1-x}Sn_x$ diamond-lattice is reached between x = 13.6% and x = 15% The incorporation of 15% Sn leads to the observed columnar defects which decrease the layer quality dramatically, even though one can still find high-quality, but partly relaxed $Ge_{1-x}Sn_x$ between these defects. These results are consistent with the loss of pendellösungs fringes in the XRD experiment, which occurs in the same concentration range, as shown in Fig. S2.1.

**Supplementary Material S3 - Temperature Stability during Growth at 300°C**

In sample series B (Table S1), we investigated the thermal stability of $Ge_{0.9}Sn_{0.1}$ layers during growth at systematically increasing growth temperatures. To identify the droplet-like surface features observed in AFM images, we prepared a FIB-cut through the sample grown at 300°C (B6) and performed quantitative TEM analyses. Fig. S3(a) displays a cross-sectional SEM image that was recorded during the preparation of the TEM lamella. From EDXS experiments and HRTEM images, we identified the islands at the surface (bright protrusions in the image) as β-Sn precipitates. The islands can be associated with solidified Sn droplets and terminate short braided trails that are visible in the SEM images in Fig. S3(b). The Sn precipitates reach down to the $Ge_{1-x}Sn_x$/Ge interface, as can be seen in Fig. S3(c).

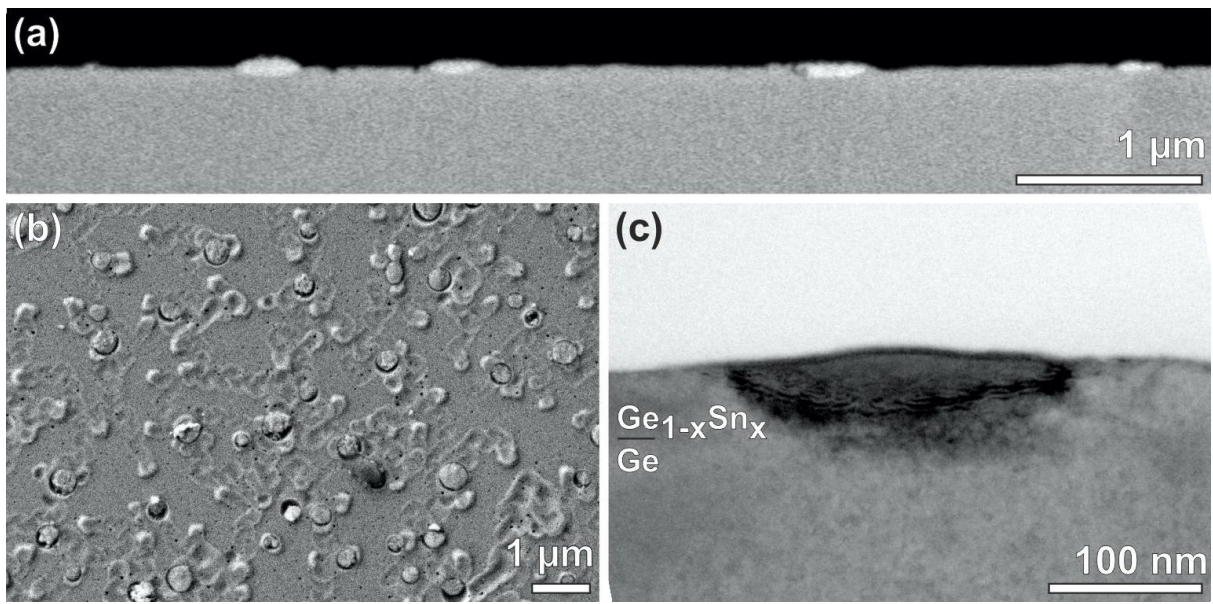

**Fig. S3**: (a) Cross-sectional SEM image of a FIB-cut lamella for TEM investigations. The bright protrusions at the surface are Sn precipitates that can be distinguished from the Ge bulk by the material contrast. (b) SEM image of the sample surface where Sn droplets decorate the end of braided trails. (c) Cross-sectional TEM image of a β-Sn precipitate.

**Supplementary Material S4 – Temperature Stability during *in-situ* Annealing**

A similar behavior as in section S3 was found for the samples of the annealing-series C. Annealing below the eutectic temperature of $Ge_{1-x}Sn_x$ did not affect the pseudomorphic epilayer. Annealing above the melting temperature leads to a decomposition of the GeSn layer into a crystalline, Ge-rich film and molten β-Sn precipitates.

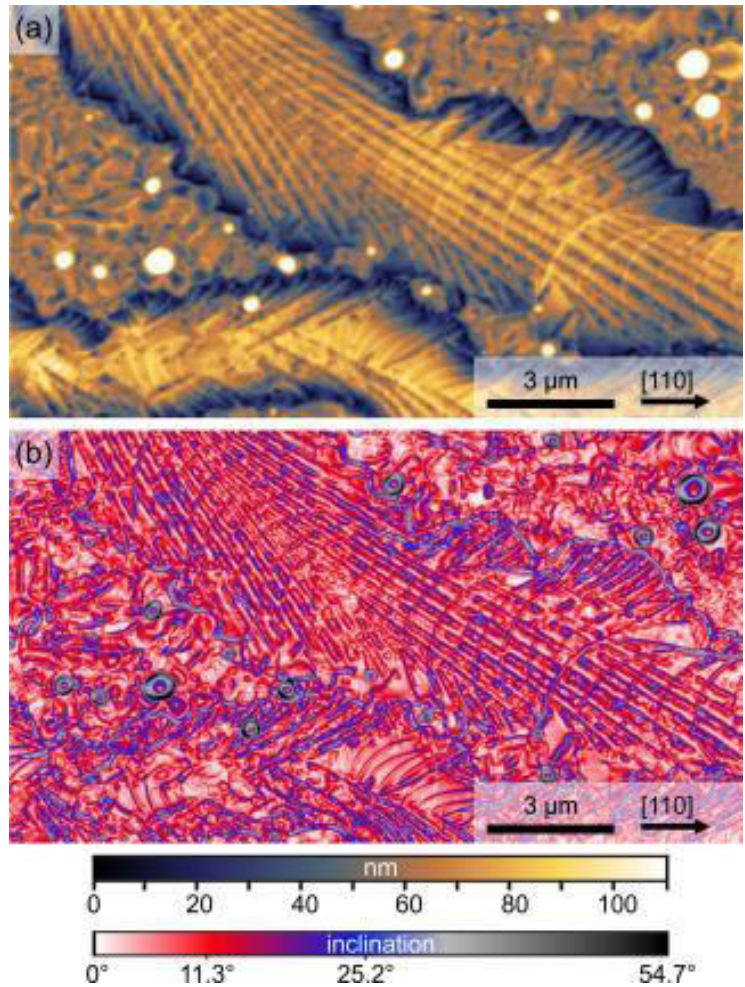

**Fig. S4**: AFM images of sample C4 which was in-situ annealed at 300°C. (a) Height image showing along the main diagonal of the image a wide Ge trail with its characteristic herring-bone pattern and the distinctive trenches that terminate the trails laterally. (b) Same image displayed as a surface-angle plot indicating the local inclinations with respect to the (001) substrate surface. The color scale is chosen in a way that highlights the predominance of {001} (0° inclination), {105} (11.3° inclination) and {113} (25.2° inclination) facets which are known to be low-energy facets of Ge.[18]

The liquid Sn droplets move over the sample surface, collect the Sn content of the layer in the contact area and leave behind crystalline, but corrugated Ge-rich trails that reveal the trajectory of droplet movement. Fig. S4(a) presents an AFM image of such a trail on sample C4. The center part of the trail has essentially the same thick as the surrounding layer, whereas the boundary on either side of the trail is depressed by about 60 nm. Overall, the missing material in the trails corresponds to good approximation to the original Sn concentration of 10% in the film, which becomes accumulated in the droplets. In Fig. S4(b) the same area is displayed as a surface-angle plot,[10] where the local inclination angles are color coded. The dominant low-energy facets of (strained) Ge[17], namely {001}, {105} and {113}, are plotted in white, red and blue to guide the eye.

**Supplementary Material S5 –Post-Growth SEM-annealing: Stop-Motion Video Sequences**

Stop-motion video sequences of our *in-situ* SEM-annealing experiments are available as video files. Sequences V1-V3 were recorded near $T_E$ at 250±25°C, V4 at 350±25°C. Video sequences V1 – V3 were simultaneously recorded with an Everhart-Thornley detector (labeled SE2 in the movies) and a through-the-lens detector[9] (labeled inLens in the movies). Video sequence V4 is only available in the inLens configuration.

The inLens SEM images show high-resolution topography contrast because mainly the primary secondary electrons generated by the incident electron beam are detected. Topography resolution is worse in the SE2 SEM images because contributions of secondary electrons induced by back scattered electrons contribute to the detected electron intensity and lead to a blurring of small topography features. Bright regions in SE2 SEM images are observed, e.g. at droplets, if the surface is inclined towards the Everhart-Thornley detector. An interesting feature is found in inLens SEM images (Fig. S5) where dark regions can be recognized on the Sn droplet or in trail behind the droplets. We associate these intensity changes to a locally higher work function which lowers the intensity of the emitted secondary electrons. Work-function changes can be, e.g., associated with compositional changes in the top few atom layers close to the surface.

The following files are available:

**Video sequence V1**: At the beginning, the video sequence shows three large Sn droplets moving downward on parallel <110> trajectories. Two of them come to a halt when encountering areas that have already been converted from the original GeSn film into re-deposited Ge by secondary droplets launched from the trails of other droplets. The smaller trails of the secondary droplets lead to an avalanche-like broadening of the converted area in the wake of the main trajectory. The trails are well resolved in the inLens video sequence, whereas the corresponding Sn droplets at the trails' ends can only be well recognized in the SE2 movie. The center one of the three original droplets runs furthest, but it is finally stopped by the trail of a fourth droplet that crosses its trajectory under an angle of 45°.

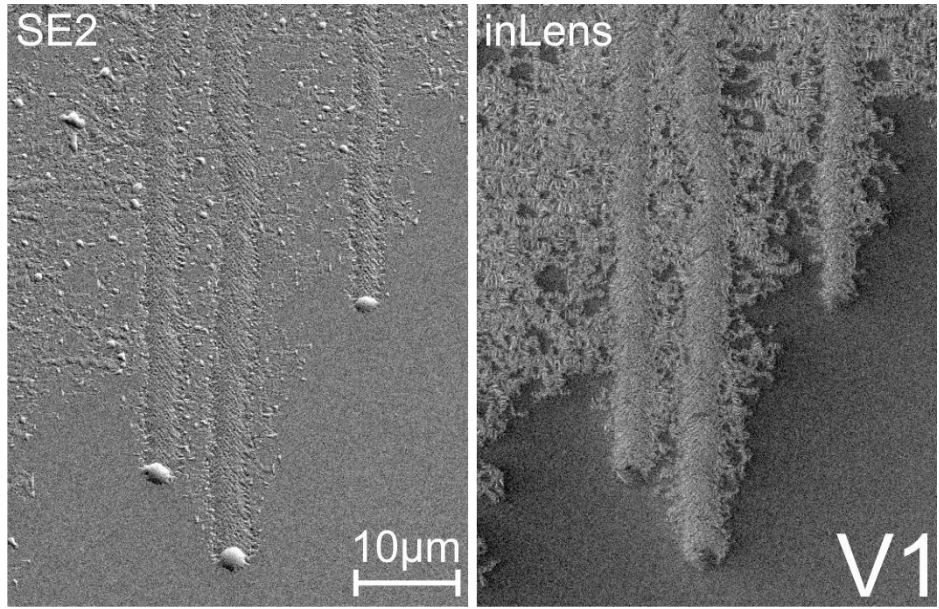

**Fig. S5.1:** Start position of video V1, which was recorded at 250°C.

Toward the end of the sequence, this fourth droplet turns in counter-clockwise in order to avoid other trails, but finally comes to a halt when it is completely surrounded by re-deposited Ge. This situation is also depicted in extracted still images that are shown in Figs. 2(d), (e) and (f) of the main text. Video V1 covers a time span of 27 min and 46s and contains 79 frames, i.e. the average time between frames is ~21s.

**Video sequence V2**: A single Sn droplet moves until it reaches already phase separated material. The video sequence shows a time span of 11 min and 4s and contains 33 frames (time between frames ~20s)

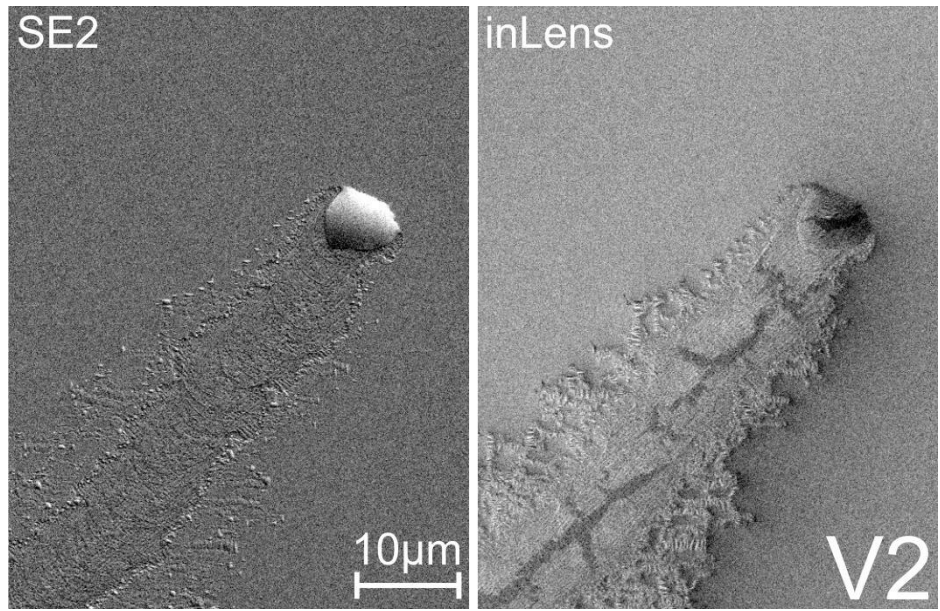

**Fig. S5.2:** Start position of video V2, which was recorded at 250°C.

**Video sequence V3**: A large Sn droplet is moving in a <100> direction in close vicinity to already transformed material until it finally comes to a halt. Note that in this less frequently observed direction of movement the leading edge of the droplet is completely facetted in a saw tooth pattern to minimize the interface energy. The video sequence shows a time span of 9 min and 1s and contains 24 frames (time between frames ~26s)

**Video sequence V4**: This inLens video sequence shows an overview of the transformation front between the smooth GeSn film in the lower left part of the field of view, and the corrugated area of re-deposited Ge left behind by a large number of moving Sn droplets. Only the trails are resolved here, whereas the droplets at their end are hardly visible with the inLens detector. This video sequence clearly shows that the avalanche-like cascade of droplets converts essentially the whole GeSn film into a corrugated, single crystalline Ge layer. Only small patches in the corrugated area remain seemingly unaffected. One can, however, observe that some of these decompose by the delayed formation of additional secondary droplets after the main transformation front has passed. Video sequence V4 provides particularly impressive evidence for the efficiency of the phase separation process induced by free-running Sn precipitates at a still moderate temperature of 350°C. The video covers a time span of 14 min and 43s and contains 37 frames (time between frames ~24s)

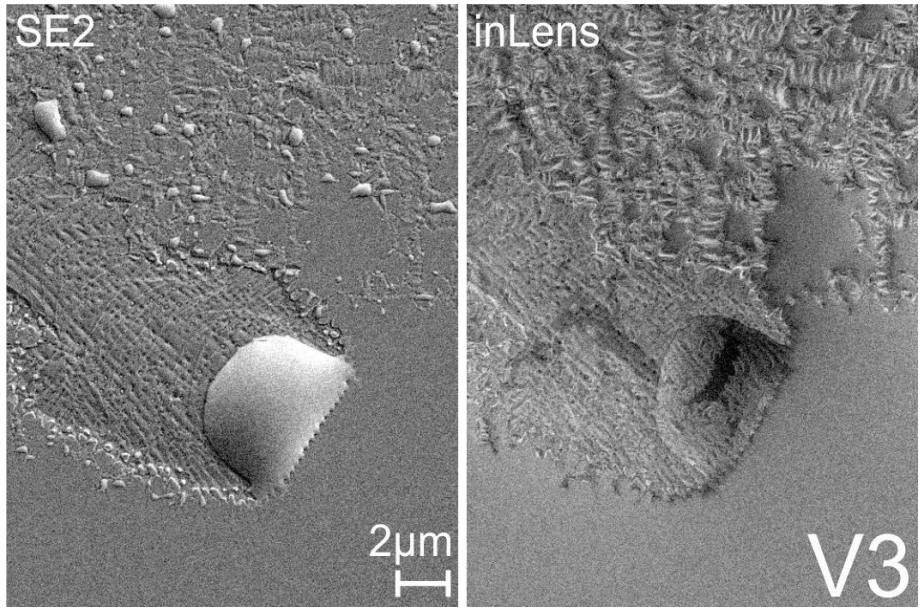

**Fig. S5.3:** Start position of video V3, which was recorded at 250°C.

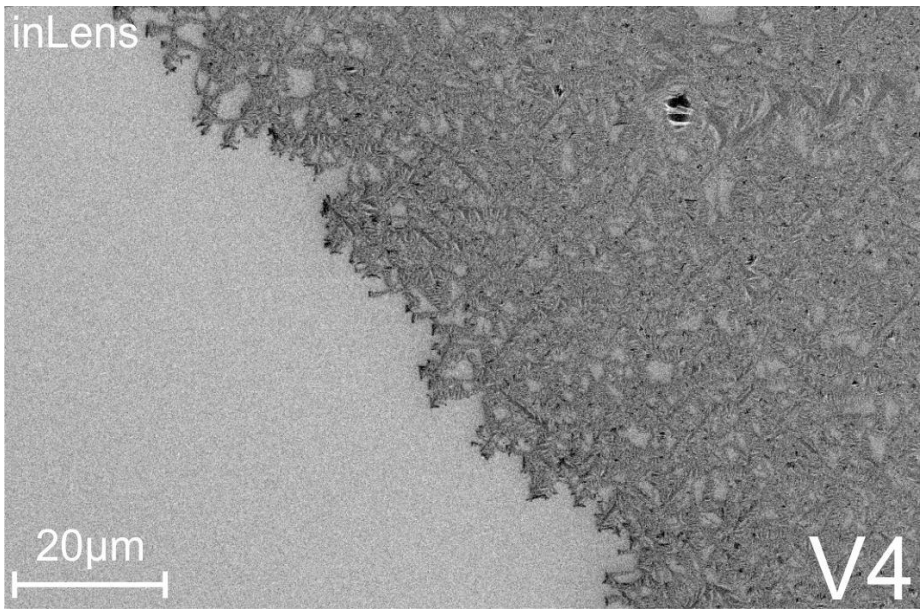

**Fig. S5.4:** Start position of video V4, which was recorded at 350°C.

**Supplementary Material S6 – Precipitate Orientation after Cool-Down**

We prepared FIB lamellae through a large Sn dot of an annealed sample from Series D after cool-down to identify the crystal structures and orientation relations of the different phases in the solid state. The droplet reaches down to the original Ge buffer/SnGe interface and crystallizes in the β-Sn crystal structure, as determined by HRTEM imaging. Due to the temperature behavior and the melting temperature of Sn, it is clear that the droplets are liquid during annealing cycles above 230°C. After cool-down, the solidified Sn droplet is surrounded by a collar of roughly triangular cross section (Fig. S5(a) and Fig. 2(b) in the main text), which we identified by EDXS as being pure Ge. Most likely, the collar consists of Ge that was originally dissolved in the liquid Sn droplet but has precipitated during the liquid-solid phase transition of the droplet. Such a behavior is consistent with the phase diagram of Ge-Sn, which has a eutectic point very close to pure Sn at a temperature about 1K below the melting point of β-Sn.[18] At temperatures above the eutectic temperature, liquid Sn can take up significant amounts of Ge which have to precipitate as almost pure Ge when the temperature falls below the eutectic temperature. An estimate of the Ge content in the melt based on the relative volume of the Ge collar is given in the main text.

To determine the alignment between the diamond lattice of the Ge film and the tetragonal lattice of the solidified β-Sn, we recorded HRTEM images from different regions of Fig. S5(a) along the [110] zone axis of the Ge substrate. Conversions by FFT into reciprocal space images are displayed in Figs. S5 (c) – (e). The assignment of the diffraction spots refers to the respective crystal structure, with blue labeling being used for β-Sn, green one for Ge. The solidified droplet itself has the β-Sn structure exhibiting the $[0\bar{1}3]$ zone axis. The β-Sn crystal lattice is oriented such that its (200) planes are almost parallel to the $(1\bar{1}\bar{1})$ planes of the diamond lattice of Ge. A schematic representation of the relative alignment of the two lattices is depicted in Fig. S5(b). This type of alignment is not perfect, as can be seen in Fig. S5(e), were the $200_{Sn}$ (blue) and the $1\bar{1}\bar{1}_{Ge}$ (green) spots do not perfectly coincide.

Fig. S6(a) also shows a thin layer containing α–GeSn nanocrystals that decorate the interface between the separated Sn and Ge phases. We assume that these have formed in the initial stages of crystallization of the Sn droplet. The α–GeSn nanocrystals can be compared to the GeSn material with high Sn-content found in capped GeSn-layers after annealing[19]. Also, the β-tin drop has developed low-energy facets during solidification, which can be seen in the SEM image of Fig 2(b) in the main text and in the asymmetric polygon-contour of the droplet in Fig. S6(a).

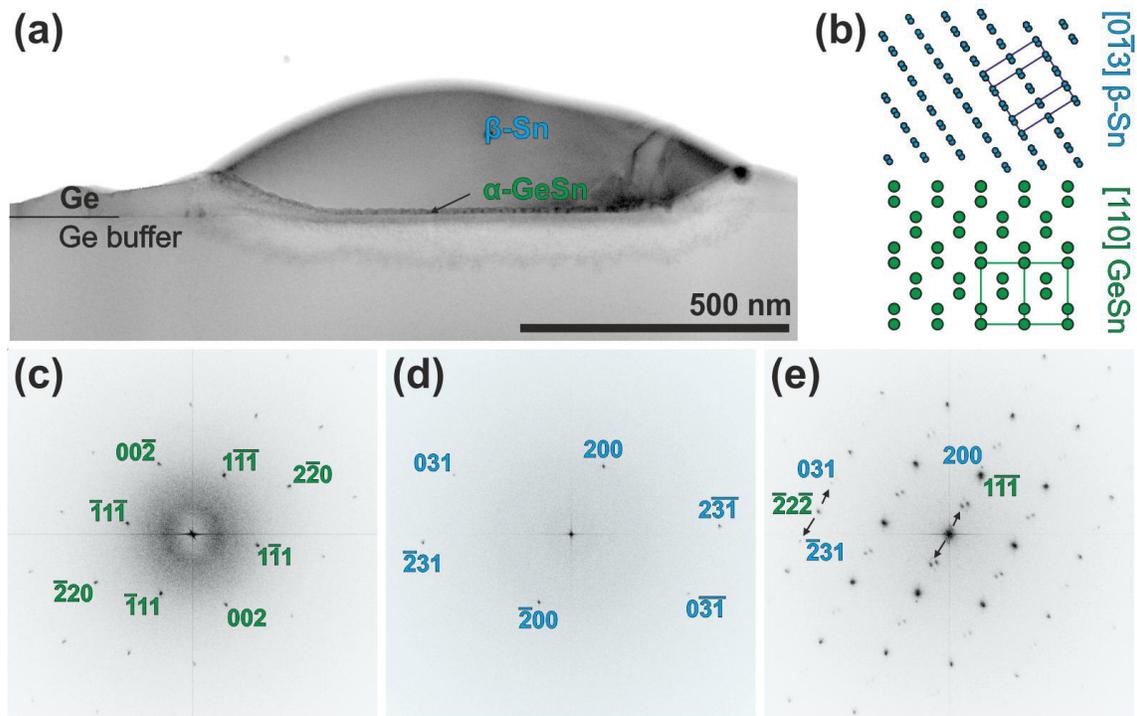

**Fig S6**: (a) Cross sectional TEM image of a large solidified Sn droplet which penetrates the whole thickness of the GeSn film down to the Ge buffer. (b) Crystal orientation of the β-Sn droplet in relation to the Ge substrate. Reciprocal space images in (c) – (e) were calculated by FFT from HRTEM images covering (c) the Ge buffer region, (d) the β-Sn droplet and (e) the interface region containing the Ge buffer, the droplet and a layer of α-GeSn precipitates that crystallize at the interface. The diffraction peaks are labeled in green for Ge signals, and in blue for Sn signals.

**Supplementary Material S7 – Decomposition Thermodynamics**[20]

One can write the Gibbs free energy $G^{GeSn}$ of Ge$_{1-x}$Sn$_x$ as the weighted sum of the Gibbs free energies $G^{Ge}$ for a Ge fraction $(1-x)$ and $G^{\alpha-Sn}$ for a α-Sn fraction $x$ as

$$G^{GeSn} = (1-x)G^{Ge} + xG^{\alpha-Sn} + \Delta G_{mix}^{GeSn}(x), \qquad \text{equ. S1}$$

with the mixing term

$$\Delta G_{mix}^{GeSn}(x) = \Delta H_{mix}^{GeSn}(x) - T\Delta S_{mix}^{GeSn}(x). \qquad \text{equ. S2}$$

The entropy of formation $\Delta S_{mix}^{GeSn}$ lowers the energy for a solution of two components and leads to an ideal solution with no miscibility gap. Ge$_{1-x}$Sn$_x$, however, exhibits a large miscibility gap. Thus, over a large composition range the heat of formation $\Delta H_{mix}^{GeSn}$ must be much larger than $T\Delta S_{mix}^{GeSn}$. Therefore $T\Delta S_{mix}^{GeSn}$ can be neglected over almost the entire composition range, resulting in

$$\Delta G_{mix}^{GeSn} \sim \Delta H_{mix}^{GeSn}. \qquad \text{equ. S3}$$

After decomposition the material is separated into almost pure, re-deposited Ge with a Gibbs free energy $G^{Ge}$ and liquid Sn with a concentration $y$ of solved Ge. The total Gibbs energy of the constituents can be written as:

$$G^{Ge+Sn(l)} = (1-x-y)G^{Ge} + xG^{Sn(l)} + yG^{Ge(l)} + \Delta G_{mix}^{Sn(l)Ge(l)} + \gamma \qquad \text{equ. S4}$$

The first term is for the re-deposited Ge (the < 1% of Sn solved in the crystalline Ge phase is neglected). The second, third and fourth terms describe liquid Sn ($xG^{Sn(l)}$) containing solved Ge ($yG^{Ge(l)}$) and the mixing term ($\Delta G_{mix}^{Sn(l)Ge(l)}$). $\gamma$ is the energy of a general interface between crystalline Ge and liquid Sn. This term will be neglected in the following, because the interface terms are contained in our liquid-phase-epitaxy model (equs. 1 and 3 in the main text). The energy gain associated with the phase separation into Ge and liquid Sn is the difference between equ. S4 and S1:

$$\Delta G^{Ge+Sn(l)} = G^{Ge+Sn(l)} - G^{GeSn} = y(G^{Ge(l)} - G^{Ge}) + \Delta G_{mix}^{Sn(l)Ge(l)} + x(G^{Sn(l)} - G^{\alpha-Sn}) - \Delta G_{mix}^{GeSn} = \Delta G_l^{Ge} + \Delta G_l^{Sn} \qquad \text{equ. S5}$$

The fraction of solved Ge in liquid Sn is determined by a trade-of between dissolving Ge in the liquid Sn

$$\Delta G_l^{Ge} = y\big(G^{Ge(l)} - G^{Ge}\big) + \Delta G_{mix}^{Sn(l)Ge(l)} > 0 \qquad \text{equ. S6}$$

and forming liquid Sn from a solid α-Sn component releasing the mixing enthalpy $\Delta H_{mix}^{GeSn}$:

$$\Delta G_l^{Sn} = x\big(G^{Sn(l)} - G^{\alpha-Sn}\big) - \Delta H_{mix}^{GeSn} < 0 \qquad \text{equ. S7}$$

$G^{Ge(l)} - G^{Ge}$ is positive, because we are far below the melting point of Ge. We have a supersaturated solution, thus $\Delta G_{mix}^{Sn(l)Ge(l)}$ is also positive, resulting in $\Delta G_l^{Ge} > 0$. Because we are near the melting point of Sn the term $x\big(G^{Sn(l)} - G^{\alpha-Sn}\big)$ should be mainly govern by the enthalpy of fusion $H_{fus}^{Sn}$ and is in the order of $xH_{fus}^{Sn} \approx 0.7 \frac{\text{kJ}}{\text{mol}}$ for $x = 0.1$.

The mixing enthalpy $\Delta H_{mix}^{GeSn}$ can be dominated by two main terms: the difference in bond energies between Ge-Ge, Sn-Sn and Ge-Sn bonds and the deformation energy induced by atomic size or bond-length mismatch of the constituents. Several publications[21, 19] indicate that the bond energies differences are small. $Sn_{1-y}Ge_y$ melts, for instance, have a minimum in the surface tension, and it is possible to fabricate regular $Ge_{0.5}Sn_{0.5}$ alloys. Both results lead to the conclusion that Ge-Sn bonds are energetically not unfavorable enough to cause the large miscibility gap. It is therefore more likely that the deformation of the Ge-matrix by the much larger Sn-atoms is the dominating term leading to the large $\Delta H_{mix}^{GeSn}$. For our system, grown far away from equilibrium, the inner energy and thus $\Delta H_{mix}^{GeSn}$ is of course also increased by e.g. defects, unsaturated bonds, Sn or Ge interstitials, etc.. We therefore assume $\Delta H_{mix}^{GeSn} \gg x\big(G^{Sn(l)} - G^{\alpha-Sn}\big)$. Thus, we can write for equ. S7 $\Delta G_l^{Sn} \sim -\Delta H_{mix}^{GeSn}$ which leads to equ. 3 in the main text:

$$\Delta G^{Ge+Sn(l)} = \Delta G_l^{Ge} - \Delta H_{mix}^{GeSn} \qquad \text{equ. S8}$$